\documentclass[twocolumn,floatfix,showpacs,superscriptaddress]{revtex4}

\usepackage{graphicx}
\usepackage{dcolumn}
\usepackage{bm}
\usepackage{amsmath}

\begin{document}
\title{Signatures of topology in ballistic bulk transport of HgTe quantum wells}

\author{E. G. Novik}
\affiliation{Physikalisches Institut (EP3), University of W\"urzburg, 97074 W\"urzburg, Germany}
\author{P. Recher}
\affiliation{Institut f{\"u}r Theoretische Physik und Astrophysik, University of W\"urzburg, 97074 W\"urzburg, Germany}
\author{E. M. Hankiewicz}
\affiliation{Institut f{\"u}r Theoretische Physik und Astrophysik, University of W\"urzburg, 97074 W\"urzburg, Germany}
\author{B. Trauzettel}
\affiliation{Institut f{\"u}r Theoretische Physik und Astrophysik, University of W\"urzburg, 97074 W\"urzburg, Germany}

\begin{abstract}
We calculate bulk transport properties of two-dimensional topological insulators based on HgTe quantum wells in the ballistic regime. Interestingly, we find that the conductance and the shot noise are distinctively different for the so-called normal regime (the topologically trivial case) and the so-called inverted regime (the topologically non-trivial case). Thus, it is possible to verify the topological order of a two-dimensional topological insulator not only via observable edge properties but also via observable bulk properties. This is important because we show that under certain conditions the bulk contribution can dominate the edge contribution which makes it essential to fully understand the former for the interpretation of future experiments in clean samples.
\end{abstract}

\date{\today}
\narrowtext\maketitle

The physics of topological insulators, that are bulk insulators with certain topological properties, is one of the most active areas in modern condensed matter research. These insulators can be either characterized by bulk Chern numbers \cite{Qi2006} or by a so-called $Z_2$ topological order, in which a system, that is invariant under time-reversal symmetry (TRS), is classified into two classes according to whether there are an even or odd number of Kramers partners of edge states at a given boundary of the system \cite{Kane2005}. The latter classification makes it illustrative how to distinguish a topologically trivial from a topologically non-trivial insulator with respect to TRS. If there is an odd number of Kramers partners at a given edge then the system is considered to be topologically non-trivial because no scattering potential that preserves TRS can scatter a left-mover into a right-mover. This scattering process is strictly forbidden by TRS \cite{Wu2006}. If instead there is an even number of Kramers partners at a given edge, the system is called topologically trivial because the edge states are not protected anymore against potential scattering by TRS and, hence, a left-mover can rather easily scatter into a right-mover and vice versa.

Only one year after the prediction has been made that HgTe quantum wells (QWs) are prime candidates for two-dimensional topological insulators \cite{BernevigScience06}, experimental evidence based on edge state transport has been found \cite{Koenig07}. In HgTe QWs, the thickness of the well controls the topology, meaning that thinner wells (below a critical thickness) are topologically trivial insulators and thicker wells (above a critical thickness) are topologically non-trivial insulators. The former is called {\it normal regime}, the latter {\it inverted regime}, and the critical thickness has been coined {\it mass-inversion point} with respect to the effective model discussed in more detail below. To the best of our knowledge, all experimental evidence both in two-dimensional \cite{Koenig07} as well as three-dimensional topological insulators \cite{Hsieh2008} is based on the physics of the edges. In this Letter, we propose a way to experimentally distinguish a trivial from a non-trivial two-dimensional topological insulator using bulk properties only. We calculate the linear conductance as well as the Fano factor (the ratio of the shot noise and the average current) for ballistic HgTe QWs focusing on bulk transport. This transport is mediated by macroscopic quantum tunneling through the bulk gap. In the topologically trivial case, where edge current does not exist at all, our predictions are directly observable, whereas in the topologically non-trivial case, the bulk transport that we calculate supplements the edge current. We show that the bulk transport in the two regimes (normal {\it vs.} inverted) looks qualitatively different such that it becomes obvious how to distinguish a trivial from a non-trivial case. Furthermore, under certain conditions which we specify below, the bulk transport can dominate the edge transport in the inverted regime. As far as the current noise is concerned, the bulk transport gives the only contribution because the edge current is noiseless. Our findings are certainly experimentally relevant and should be taken into account when analyzing ballistic transport properties of HgTe nanostructures.

\begin{figure}[h]
\begin{center}
\includegraphics[width=0.75\columnwidth]{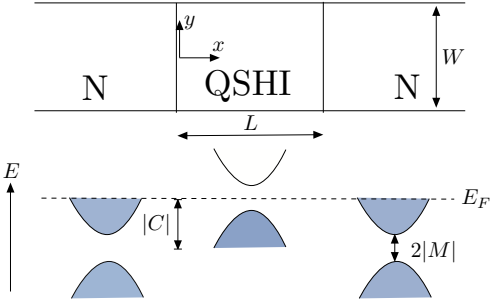}
\end{center}
\caption{(Color online) Schematic of the transport setup. Normal metal reservoirs (N) are attached to a ballistic quantum spin Hall insulator (QSHI) of width $W$ and length $L$. The metal reservoirs are modeled as highly doped HgTe as shown schematically in the lower part of the figure.}
\label{fig_setup}
\end{figure}

In order to describe a ballistic HgTe QW coupled to normal metal leads, we use the effective Hamiltonian, derived in Ref.~\cite{BernevigScience06},
\begin{equation}
\label{H1}
H=\left(\begin{array}{cc}h(k) &  0 \\ 0 & h^{*}(-k)\end{array}\right)
\end{equation}
with $h(k)=\epsilon(k){\rm I}_{2\times2}+d_{a}(k)\sigma^{a}$ and
$d_{a}(k)=(Ak_{x},-Ak_{y},M(k))$ where $\sigma^{a}$ is the vector
of Pauli matrices. In Eq.~(\ref{H1}), $\epsilon(k)=C-Dk^2$, and the mass
term $M(k)=M-Bk^2$ with $k^2=k_x^2+k_y^2$. The parameters
$A,B,C,D,M$ depend on the QW geometry. Realistic estimates can be made by a comparison of the effective model with a well established $8 \times 8$ Kane Hamiltonian \cite{Schmidt}. In the transport situation, we set $C$ to zero in
the center region, which we call quantum spin Hall insulator (QSHI), and to a large negative value in the leads, thereby effectively modeling the
metal contacts by highly doped HgTe, {\it {cf.}} Fig.~\ref{fig_setup} for a schematic. We note that our model for metallic
leads is similar to the corresponding model in graphene \cite{Tworzydlo} which has already been successfully applied to the interpretation of ballistic transport in graphene nanostructures, see, for instance, the shot noise measurement in Ref.~\cite{Dan08}. It differs however from the model introduced in Ref.~\cite{Nagaosa} where the metal contacts have a higher degeneracy of states than the QSHI region. Our model avoids this additional degeneracy which we find the more realistic scenario.

The energy eigenvalues of $h(k)$ for constant $C$ are given by $E_{\pm}=C-Dk^2\pm d(k)$ with $d(k)=\sqrt{(Ak)^2+M^{2} (k)}$ where
$\pm$ denote solutions of the two bands (separated by an energy gap $2|M|$ at $k=0$). Since the transmission coefficients for the two blocks $h(k)$ and $h^{*}(-k)$ of the Hamiltonian (\ref{H1}) are degenerate in presence of TRS, we restrict ourselves to $h(k)$ in the following discussion but take the degeneracy into account in all final results.

To calculate transport through the system we need to match the scattering states $\psi^{n}_{E}(x,y)$ and their associated
currents $(\partial h(k)/\partial k_{x})\psi^{n}_{E}(x,y)$ (for the three different regions in space N-QSHI-N) at the crossover points $x=0$ and $x=L$. This has to be done at fixed energy $E$ and mode index $n$. In order to avoid spurious edge effects, we choose periodic (PBC) and antiperiodic boundary conditions (APBC) in transverse direction, yielding a quantization of the transverse momentum $k_y^{n}$ via $k_y^{n}=2\pi n/W$ (PBC) and $k_y^{n}=\pi (2n+1)/W$ (APBC) with mode index $n=0,\pm 1,\pm 2,\cdots$. The two types of boundary conditions are then compared to identify how a particular result depends on the choice of boundary conditions. In a finite system (with edges), the more appropriate boundary conditions are hard-wall ones as discussed in Ref.~\cite{Zhou08}. However, hard-wall boundary conditions imply edge states which would mask the bulk properties discussed here. In practice, our predictions will add up to the edge state contribution to the current. Furthermore, it is reasonable to expect that, in the regime $W/L \gg 1$, all observable quantities will not depend on a particular choice of boundary conditions anymore. We now discuss the explicit expressions of the scattering problem for positive energy solutions. The scattering states have the form
$\psi_{E}^{n}(x,y)=e^{ik_{1}\sin\theta_{1}^{n}y}\phi_{E}^{n}(x)$ with $k_{1}\sin\theta_{1}^{n}=k_{y}^{n}$ where $\theta_{1}^{n}$ is the angle of incidence of mode $n$. For the incoming scattering state in the left reservoir ($x<0$), we make the ansatz
\begin{eqnarray} \label{phi1}
\phi_{E}^{n}(x)&=&\left(\begin{array}{c}Ak_{1}e^{i\theta_{1}^{n}}\\d(k_{1})-M(k_{1})\end{array}\right)e^{ik_{1}\cos\theta_{1}^{n}x} \nonumber \\
&+&r_{1}^{n}\left(\begin{array}{c}-Ak_{1}e^{-i\theta_{1}^{n}}\\d(k_{1})-M(k_{1})\end{array}\right)e^{-ik_{1}\cos\theta_{1}^{n}x}\nonumber\\
&+&r_{2}^{n}\left(\begin{array}{c}-Ak_{2}e^{-i\theta_{2}^{n}}\\d(k_{2})-M(k_{2})\end{array}\right)e^{-ik_{2}\cos\theta_{2}^{n}x}
\end{eqnarray}
with the mode-dependent reflection coefficients $r_{1}^{n}$ and $r_{2}^{n}$. We have defined the wave vectors at the Fermi energy $E_F$ as
\begin{equation} \label{k12}
k_{1,2}=\frac{\sqrt{\Delta\pm\sqrt{\Delta^2-4(B^2-D^2)(M^2-\tilde{C}^2)}}}{\sqrt{2(B^2-D^2)}} ,
\end{equation}
where $\Delta=-A^2+2MB-2\tilde{C}D$ and $\tilde{C}=C-E_F$ with the choice $|B|>|D|$ such that $k_{1}$ is real. Note that $k_2=i|k_{2}|$ is purely imaginary and therefore describes evanescent waves in Eq.~(\ref{phi1}). The outgoing
scattering state, in the right reservoir ($x>L$), has the form
\begin{eqnarray}
\phi_{E}^{n}(x)&=&t_{1}^{n}\left(\begin{array}{c}Ak_{1}e^{i\theta_{1}^{n}}\\d(k_{1})-M(k_{1})\end{array}\right)e^{ik_{1}\cos\theta_{1}^{n}x}\nonumber\\
&+&t_{2}^{n}\left(\begin{array}{c}Ak_{2}e^{i\theta_{2}^{n}}\\d(k_{2})-M(k_{2})\end{array}\right)e^{ik_{2}\cos\theta_{2}^{n}x}
\end{eqnarray}
with transmission coefficients $t_{1}^{n}$ and $t_{2}^{n}$.
Within the QSHI, $0\leq x\leq L$, the wave
function is
\begin{eqnarray}
\phi_{E}^{n}(x)&=&\alpha_{3}^{n}\left(\begin{array}{c}Ak_{3}e^{i\theta_{3}^{n}}\\d(k_{3})-M(k_{3})\end{array}\right)e^{ik_{3}\cos\theta_{3}^{n}x}\nonumber\\
&+&\alpha_{4}^{n}\left(\begin{array}{c}Ak_{4}e^{i\theta_{4}^{n}}\\d(k_{4})-M(k_{4})\end{array}\right)e^{ik_{4}\cos\theta_{4}^{n}x}\nonumber\\
&+&\beta_{3}^{n}\left(\begin{array}{c}-Ak_{3}e^{-i\theta_{3}^{n}}\\d(k_{3})-M(k_{3})\end{array}\right)e^{-ik_{3}\cos\theta_{3}^{n}x}\nonumber\\
&+&\beta_{4}^{n}\left(\begin{array}{c}-Ak_{4}e^{-i\theta_{4}^{n}}\\d(k_{4})-M(k_{4})\end{array}\right)e^{-ik_{4}\cos\theta_{4}^{n}x},
\end{eqnarray}
where $k_{3,4}$ are given by the same expressions as $k_{1,2}$ (see Eq.~(\ref{k12})) with $C=0$ and the coefficients $\alpha_{3}^{n}$, $\alpha_{4}^{n}$, $\beta_{3}^{n}$, and $\beta_{4}^{n}$ determine the weight of the different parts of the wave function in the center region. Note that $k_{3,4}$ are purely imaginary in the gap (for $E_F=0$) for typical parameters of realistic HgTe QWs.

The conductance $G$ and Fano factor $F$ of bulk transport are then obtained using Landauer transport theory via $G=(2e^2/h)\sum_{n=-N}^{N}T_{1}^{n}$ and $F=\sum_{n=-N}^{N}T_{1}^{n}(1-T_{1}^{n})/\sum_{n=-N}^{N}T_{1}^{n}$, where $T_{1}^{n}\equiv |t_{1}^{n}|^2$ and $N$ is the number of propagating modes in the leads, i.e. the largest integer such that $k_{y}^{n}<k_{1}$.

In the following, we discuss in detail the dependence of the conductance and the Fano factor on different parameters of the effective Hamiltonian (\ref{H1}). It is well-known that the sign of $M/B$ matters for the classification of the topological order of the system \cite{Zhou08}. If $M$ and $B$ have the opposite sign, the system is in the normal regime, while, if they have the same sign, it is in the inverted regime. We will show that for samples with realistic aspect ratios, the measurement of either bulk conductance or shot noise is sufficient to uniquely identify the topological order of the HgTe QW.
%
\begin{figure}[h]
\begin{center}
\includegraphics[width=0.7\columnwidth]{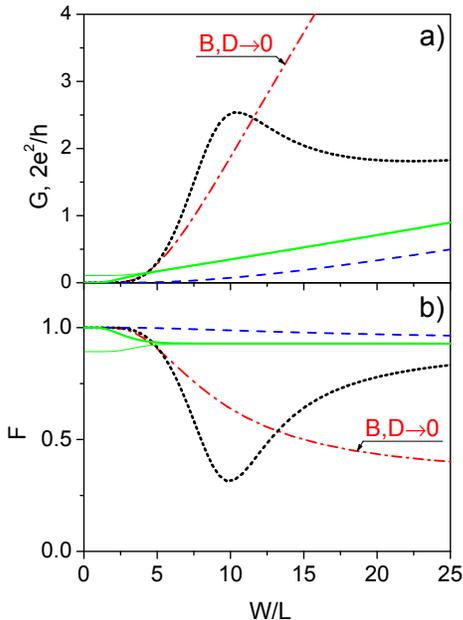}
\end{center}
\caption{(Color online) (a) The conductance $G$ and (b) the Fano factor $F$ as a function
of aspect ratio $W/L$  for parameters typical for HgTe quantum wells $B=-1.120~{\rm eV~nm}^2$, $D=-730~{\rm meV~nm}^2$ under the choice $E_{F}=0$ (corresponding to bulk transport through the gap). Metallic leads are modeled by $C=-3.75~{\rm eV}$. The solid (green) lines correspond to the case $M \rightarrow 0$, the dashed (blue) lines to $M>0$ ($M \approx 3.1~{\rm meV}$), the {\it normal regime}, and the dotted (black) lines to $M<0$ ($M \approx -3.1~{\rm meV}$), the {\it inverted regime}. In the case $M=0$, the thin solid (green) line for PBC differs slightly from the thick solid (green) line for APBC. In all other cases, the plots for PBC and for APBC are indistinguishable.
For comparison, we also show the corresponding plot for finite $M$ in the absence of quadratic terms in the Hamiltonian ($B,D \rightarrow 0$). Then the curves for positive and negative $M$ lie on top of each other, shown as a dashed-dotted (red) line. In this situation, the Fano factor approaches the universal value $1/3$ for $W/L \gg 1$. (Note that since we fix $A$, $M$, and $W$ in the plots and decrease $L$ with increasing aspect ratio $W/L$, the limit $W/L \gg 1$ implies that $|M|L/A \ll 1$.)}
\end{figure}
%
In Fig.~2, we plot the conductance $G$ as a function of geometrical
aspect ratio $W/L$ for different parameters and different boundary conditions (PBC and APBC). Here, we fix $A=375~$meV nm and $W=1~\mu$m as typical values for HgTe QWs
in the ballistic regime and vary the other parameters.  To carefully investigate the competition between
the band gap ($M$) and the effective mass parameters ($B$ and
$D$), we tune values of $M$ from positive via zero to negative and change the values of $B$ and $D$ from very small (linear dispersion $\rightarrow$ Dirac fermions) to typical ones for HgTe quantum wells \cite{Schmidt}. Let us discuss the simpler case of Dirac fermions first. In the limit $|C| \rightarrow \infty$, it is possible to derive an analytical result for the conductance of massive Dirac fermions
\begin{equation}\label{Mass_Fermions}
G=\frac{2e^2}{h}\sum_{n=-N}^{N}\left|\frac{\cos(\theta_{3}^{n})}{i
   \cos(\varphi_3^n)\cos(\theta_{3}^{n})+\sin(\varphi_3^n)\frac{E_F}{\sqrt{E_F^2-M^2}}}\right|^2,
\end{equation}
where $N\rightarrow\infty$, $\theta_{3}^{n}=\arcsin\left(\frac{2
\pi n A}{W \sqrt{E_F^2-M^2}}\right)$ and
$\varphi_3^n=\sqrt{\frac{E_F^2-M^2}{A^2}-\left(\frac{2 \pi
n}{W}\right)^2} L$.

How do conductance and Fano factor now depend on the aspect ratio $W/L$ at $E_F=0$? On the one hand, the conductance starts from zero and increases linearly with $W/L$ for large aspect ratios, pointing towards a universal conductivity $\sigma = (L/W) G$ in the limit $W/L \gg 1$. On the other hand, the Fano factor is $1$ for small values of $W/L$ and reaches $1/3$ for $W/L \gg 1$ (if $|M|L/A \ll 1$ holds at the same time). This is all not very surprising keeping the similarity to graphene in mind \cite{Tworzydlo}.

However, the surprise comes if we now take finite values of $B$ and $D$ into account. In that situation, the dependence of conductance and Fano factor on aspect ratio is crucially affected by the sign of $M/B$. For $M/B < 0$ the conductance weakly increases and the Fano factor weakly decreases with $W/L$. For $M/B > 0$, in contrast, the conductance (and Fano factor) are non-monotonic functions of $W/L$, see, for instance, the pronounced maximum for the conductance (and minimum for the Fano factor) at around $W/L \approx 10$ in Fig.~2. Due to the complexity of the underlying equations, we do not have a closed form of the position of the extremum. Nevertheless, we can find an approximate result in the parameter regime $BC \gg A^2 \gg |M|B$ with $D=0$. It shows that the distance between the leads at which the maximum in the conductance occurs is approximately given by $L_{\rm max} \approx \frac{A}{2|M|} \ln (\frac{2BC}{A^2})$ (see Appendix for the derivation).
Fortunately, the non-monotonic behavior shown in Fig.~2 does not depend much on the choice of boundary conditions (PBC or APBC) which makes us confident that it is rather robust and, thus, observable in real systems. The bottom line is that the predicted non-monotonicity is a unique signature of the inverted regime that makes it rather simple to distinguish it from the normal regime even on a qualitative level.

It becomes clear by looking at Fig.~2 that the bulk transport can be of equal magnitude or even larger than the edge transport which is limited to a contribution of $(2e^2/h)$ to the conductance and does not contribute to the Fano factor at all (because the edge current is noiseless). Therefore, our predictions are very important for the interpretation of future experiments certainly for noise measurements but also for conductance measurements of clean samples with a small band gap and a substantial aspect ratio. However, for samples used so far in experiments \cite{Koenig07}, the aspect ratios are in a range $0.5<W/L<2$
and, thus, our analysis shows that the corresponding bulk conductance is a very small fraction of the edge
conductance $2e^2/h$ in this case.

\begin{figure}[tbh]
\begin{center}
\includegraphics[width=0.8\columnwidth]{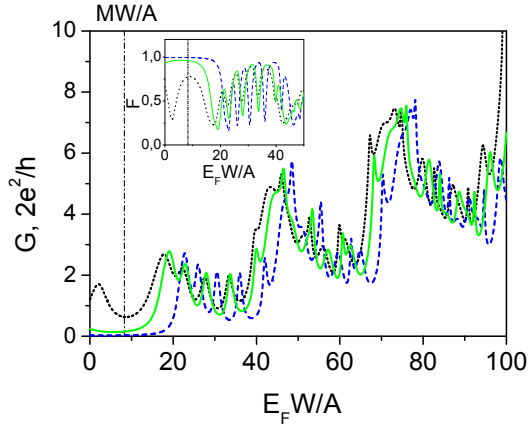}
\end{center}
\caption{(Color online) The conductance and the Fano factor (in the inset) as a function
of Fermi energy $E_{F}W/A$ for the aspect ratio $W/L=7$ and $MW/A\approx-8.3$ (dotted black lines), $MW/A\approx8.3$ (dashed blue lines) and $MW/A\rightarrow0$ (full green lines) with the choice of parameters $B=-1.120~{\rm eV~nm}^2$, $C=-3.75~{\rm eV}$, and $D=-730~{\rm meV~nm}^2$. The figure and the inset are done for PBC. Corresponding data points for APBC look similar.}
\end{figure}

Finally, in Fig.~3, we turn to the Fermi energy dependence of transport.
Interestingly, for certain aspect ratios, the conductance $G$ shows a distinct minimum for small Fermi energies $0<E_FW/A< 20$ (close to $E_F = |M|$) in the inverted regime ($M/B>0$) while it behaves qualitatively different for the normal regime ($M/B<0$) and the massless limit ($M=0$). Compare the solid, the dashed, and the dotted lines in Fig.~3. Therefore, similar to the aspect ratio dependence, the topologically non-trivial insulator can be characterized by a non-monotonic Fermi energy dependence for small Fermi energies close to the charge neutrality point. For larger Fermi energies, $G$ increases in a step-like manner corresponding to the opening of new propagating modes. On top of the step-like behavior of the conductance, one can see very pronounced Fabry-Perot oscillations in Fig.~3. In this (large Fermi energy) regime, the different types of topological order behave rather similarly and it is difficult to distinguish them from each other. In the inset, we show the Fano factor as a function of Fermi energy.
One can see that the shot noise has a Poissonian character for  small $E_F$ and $M/B \leq 0$ while it behaves
non-monotonically and has sub-Poissonian character for $M/B > 0$. This difference between the noise of the system in the normal regime and the noise of the system in the inverted regime should be easily seen in experiments and is one of the key results of our analysis. The maximum of $F$ corresponds to the minimum of $G$. For larger Fermi energy, the Fano factor has an oscillating character corresponding to the Fabry-Perot oscillations in the conductance.

In summary, we have analyzed the ballistic transport in bulk HgTe quantum wells. Our analysis of the bulk conductance complements the edge-state transport and will be important for the interpretation of future experiments in clean samples. The predicted dependence of the bulk shot noise should be directly observable because the edge current is noiseless. We have shown that both the conductance and the noise behave qualitatively different depending on whether the quantum well is in the normal or the inverted regime. This yields a new opportunity to experimentally verify the topological order of 2D topological insulators on the basis of ballistic transport measurements.

We would like to thank A. Akhmerov, H. Buhmann, L.W. Molenkamp, M.J. Schmidt, and S.C. Zhang for interesting discussions. Financial support by the German DFG is gratefully acknowledged (E.G.N. via grant no.~AS327/2-1, P.R. via the Emmy-Noether program, and E.M.H. via grant no.~HA5893/1-1.).

\appendix
\section{Analytical considerations of non-monotonic behavior of transmission}

In this Appendix, we discuss the non-monotonic behavior of the conductance with length $L$ (for fixed $W$) characteristic  for the case of a non-trivial topological insulator $(M/B>0)$ as shown in Fig.~2. To capture the essential physics with the simplest possible model, we calculate analytically the transmission coefficient for the mode $n=0$ ($T_{1}^{0}$) that corresponds to an incident
wave propagating perpendicularly to the interface between metal and
insulator (normal incidence). We consider this particular case
because the transmission probability is maximal for normal
incidence and decreases with increasing angle of incidence
rather rapidly for larger values of $L$. Furthermore, the maxima of the conductance and $T_{1}^{0}$ lie close to each other for a wide range of parameter choices. Although the behavior of $T_{1}^{0}$ for $M/B>0$ shows already the non-monotonic behavior, let us emphasize that this feature appears also for some higher modes with a small angle of incidence and is likewise present for antiperiodic boundary conditions.

We find that $T_{1}^{0}$ for both types of topological insulators ($M/B<0$ and $M/B>0$)
can be written in a simple form:
\begin{equation}
T_{1}^{0}(M/B<0)\approx\frac{4A^2 B C e^{2i\varphi_3^0}}
{4B^{2}C^{2}+A^{4}e^{2i\varphi_3^0}(2+e^{2i\varphi_3^0})},
\label{TransmM1}
\end{equation}
\begin{equation}
T_{1}^{0}(M/B>0)\approx\frac{4 A^{2} B C e^{2i\varphi_3^0}}
{4B^{2}C^{2}e^{4i\varphi_3^0}+A^{4}(1+2e^{2i\varphi_3^0})},
\label{TransmM2}
\end{equation}
where $\varphi_3^0\approx i|M|L/A$. In the latter expressions, we put $E_{F}=0$ and $D=0$, in order to obtain simple limits and use the  following set of assumptions: (i) $BC\gg A^{^2}\gg
|M|B$ which is well satisfied for typical HgTe quantum wells close to the
mass-inversion point, (ii) $e^{i\varphi_4^0}\approx e^{A L/B}\approx0$
($A>0, B<0$) which is valid for $W/L\lesssim 50$ ($W=1\mu{\rm m}$).
Note that Eqs.~(\ref{TransmM1}) and (\ref{TransmM2})
can not be applied to the case of massive Dirac fermions (the limit
$B\rightarrow0$) because it would be in contradiction with the
first assumption. However, we have recovered the limit of massive Dirac fermions when $B$ and $D$
were properly set to zero, see Eq.~(\ref{Mass_Fermions}).
\begin{figure}[h]
\begin{center}
\includegraphics[width=0.82\columnwidth]{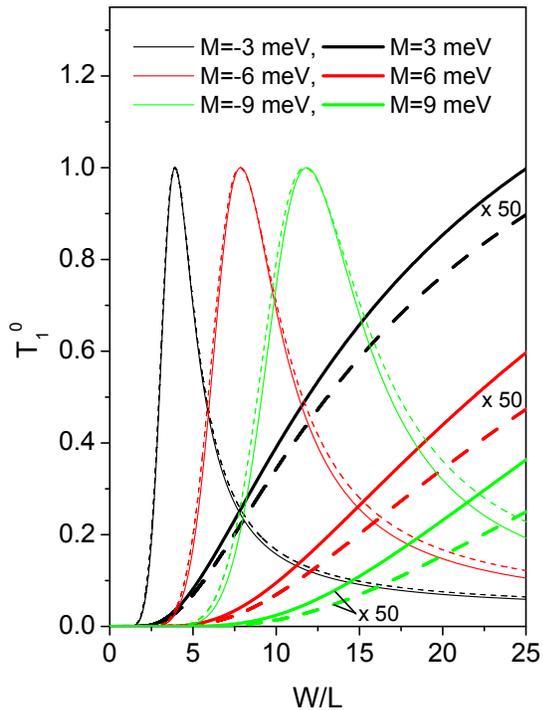}
\end{center}
\caption{(Color online) Transmission probability for normal
incidence ($n=0$). Solid lines correspond to the exact numerical
calculations, dashed lines are obtained using analytical
expressions for $T_{1}^{0}$ given in Eqs.~(\ref{TransmM1}) and (\ref{TransmM2}).
Values of $T_{1}^{0}$ for $M>0$ are  multiplied by the factor $50$ to allow for the comparison of the
$M<0$ and $M>0$ cases  on the same graph.
Here, the other parameters are chosen as $E_{F}=0$, $D=0$, $A=375~$meV
nm, $B=-1.120~{\rm eV~nm}^2$, $C=-3.75$~eV and $W=1~\mu$m .}
\label{Binfl}
\end{figure}

Eqs.~(\ref{TransmM1}) and (\ref{TransmM2}) are in a very good
agreement with the exact numerical calculations of $T_{1}^{0}$, that can be
seen from Fig.~\ref{Binfl}.  Using the assumptions given above,
the expression for $T_{1}^{0}(M/B<0)$ can be approximated further as
\begin{equation}\label{TranP}
T_{1}^{0}(M/B<0)\approx \frac{A^2}{BC}e^{2i\varphi_3^0},
\end{equation}
which means that we have exponentially decreasing transmission
probability with increasing length $L$ typical for trivial
insulators. For $M/B>0$ one should consider two limits related to
the value of $L$. In the case $e^{2i\varphi_3^0}\ll 1$ (i.e. for
relatively large values of $L$), $T_{1}^{0}$ can be approximated
as
\begin{equation}\label{TranN1}
T_{1}^{0}(M/B>0)\approx \frac{4 BC}{A^2}e^{2i\varphi_3^0},
\end{equation}
and we have the similar behavior as for $M/B<0$. In the case
$e^{2i\varphi_3^0}\sim 1$ (i.e. for relatively small values of $L$),
$T_{1}^{0}$ can be approximated by
\begin{equation}\label{TranN2}
T_{1}^{0}(M/B>0)\approx \frac{A^2}{BC}e^{-2i\varphi_3^0},
\end{equation}
and we have increasing transmission probability with increasing
length $L$ which is very unusual for a bulk insulator. The value of
$L$ corresponding to the maximum in the transmission probability
for $M/B>0$ can be found from the condition $\partial
T_{1}^{0}(M/B>0)/\partial L=0$, which gives:
\begin{equation}\label{Maximum}
L_{max}\approx\frac{A}{2|M|}\ln\left[\frac{2 B
C}{A^{2}}\right].
\end{equation}
Thus the position of the maximum depends strongly on the absolute
value of M (since the parameter $A$ changes insignificantly for typical HgTe quantum wells close
to the mass-inversion point). Indeed, the maximum of $T^0_1$ in Fig.~\ref{Binfl}, is shifted  to  the right (smaller $L$ values) when $|M|$ increases.

\end{document}